# Thickness-Controlled Black Phosphorus Tunnel Field-Effect Transistor for Low Power Switches

Seungho Kim[1], Gyuho Myeong[1], Wongil Shin[1], Hongsik Lim[1], Boram Kim[1], Taehyeok Jin[1], Sungjin Chang[2], Kenji Watanabe[3], Takashi Taniguchi[3], Sungjae Cho[1*]

The continuous transistor down-scaling has been the key to the successful development of the current information technology. However, with Moore′s law reaching its limits the development of alternative transistor architectures is urgently needed[1]. Transistors require at least 60 mV switching voltage for each 10-fold current increase, i.e. subthreshold swing (SS) 60 mV/dec. Alternative tunnel field-effect transistors (TFETs) are widely studied to achieve a sub-thermionic SS and high $I_{60}$ (current where SS becomes 60 mV/dec)[2]. Heterojunction (HJ) TFETs bear promise to deliver high $I_{60}$, but experimental results do not meet theoretical expectations due to interface problems in the HJs constructed from different materials. Here, we report a natural HJ-TFET with spatially varying layer thickness in black phosphorus (BP) without interface problems. We achieved record-low average SS over 4−5 decades of current, $SS_{ave\_4dec} \approx 22.9$ mV/dec and $SS_{ave\_5dec} \approx 26.0$ mV/dec with record-high $I_{60}$ (= 0.65−1 μA/μm), paving the way for the application in low power switches.

[1] Department of Physics, Korean Advanced Institute of Science and Technology (KAIST), Daejeon, Korea

[2] Measurement and Analysis Team, National Nanofab Center, Daejeon, Korea

[3] National Institute for Materials Science, Namiki Tsukuba Ibaraki 305-0044, Japan

Almost all aspects of life and society have changed in the past fifty years through dramatic improvements in our ability to process and communicate digital data, resulting from an integration and down-scaling of complementary metal-oxide-semiconductor (CMOS) transistors by Moore's law. However, further scaling transistors has stalled mainly due to power consumption[1]. Pop (2010)[3] showed that formerly negligible standby power consumption had reached the level of switching power consumption. Reducing both switching and standby power consumption while further scaling transistors requires overcoming the thermionic limit of SS = 60 mV/dec in conventional metal-oxide-semiconductor field-effect transistors (MOSFETs) (Supplementary Sections 1 and 2). The fundamental SS limit in the MOSFETs originates from thermal carrier injection mechanism, which prevents further reduction of a transistor supply voltage $V_{DD}$ needed to switch a transistor from off to on state. Furthermore, reduction of power consumption with SS < 60 mV/dec should be accompanied by high on-current enough to drive the subsequent transistors at fast rates. The International Roadmap for Devices and Systems predicts that new device geometries with new materials beyond CMOS will be required to address transistor scaling challenges in the near future[4]. Nikonov and Young[5] compared ultimate circuit performance for many CMOS device alternatives and identified tunnel transistors as a promising technology. In particular, TFETs have been suggested as a major alternative to MOSFETs, since SS in TFETs can be substantially reduced below the thermionic limit (60 mV/dec) due to the cold charge injection mechanism of band-to-band tunneling (BTBT). Despite intensive research[6], two critical simultaneous-requirements for TFETs to replace MOSFETs remain unfulfilled for low power applications: 1) $SS_{ave\_4\text{-}5dec}$ ($SS_{ave}$ over 4–5 decades of current) < 60 mV/dec at room temperature, and 2) high $I_{60}$ = 1–10 µA/µm, comparable to the level of the on-current at threshold voltage in state-of-the-art MOSFETs. To date, only two n-type and no p-type TFETs have been reported to achieve SS$_{ave\_4dec}$ < 60 mV/dec[7, 8] at $T$ = 300 K, but with 2–5 orders of magnitude lower $I_{60}$ than the preferred range. Increasing $I_{60}$ is essential to operate a TFET-based logic gate at a faster rate since on-current is inversely proportional to the switching delay time.

* Corresponding author, Email: sungjae.cho@kaist.ac.kr

$I_{60}$ depends strongly on BTBT transmission probability, which can be calculated by Wentzel–Kramer–Brillouin (WKB) approximation[9],

$$T_{\text{WKB}} \approx exp\left(-\frac{4\lambda\sqrt{2m^*E_g^3}}{3e\hbar(E_g+\Delta\phi)}\right), \tag{1}$$

where $\Delta\phi$ is the BTBT energy window, $\lambda$ is the screening tunneling length, $m^*$ is the carrier effective mass, and $E_g$ is the bandgap. To achieve high $I_{60}$, transmission probability should be maximized by minimizing $E_g$, $m^*$, and $\lambda$. Computational models have shown that atomically thin two-dimensional (2D) channel materials have an advantage in reducing $\lambda$ by gate modulation over 3D materials[10]. Device simulations have shown that BTBT in HJs combining small bandgap source and large bandgap channel material could significantly increase $I_{60}$ and reduce SS while keeping $I_{\text{off}}$ low[11-13]. However, interface problems due to defects, oxides, and lattice mismatches pose a major roadblock in developing high performance HJ-TFETs. Sarkar et al.[8] achieved SS$_{ave\_4dec}$ ≈ 31 mV/dec in a MoS$_2$/Ge vertical HJ-TFET. However, contrary to theoretical expectations, the HJ TFET exhibited very low $I_{60}$ ≈ 4.2×10$^{-5}$ µA/µm with low $I_{60}/I_{\text{off}}$ ≈ 7.0×10$^3$ due to the oxide tunnel barrier at the interface in the artificial HJ that formed by integrating MoS$_2$ and Ge.

We use bulk and monolayer (ML) BP 2D materials as the source and the channel of the HJ-TFETs respectively (Fig. 1a), to take advantage of unique BP band properties to solve the main problem of performance degradation for HJ-TFETs; 1) Direct bandgap changes with layer thickness from $E_g \approx 2.0$ eV in ML BP to $E_g \approx 0.3$ eV in bulk BP[14] (Supplementary Fig. 3). This thickness-dependent band property allows us to solve the major issue for HJ-TFETs. HJs can be created by varying BP thickness rather than integrating different materials, avoiding many interface problems that degrade TFET performance. 2) Effective carrier mass = 0.15 and 0.17 $m_e$ (hole and electron, respectively) along the armchair direction (Supplementary Fig. 4), which is much lighter than other 2D materials with nonzero bandgap, including transition metal dichalcogenides (MoS$_2$: 0.55–0.56 $m_e$, MoSe$_2$: 0.49–0.61 $m_e$, and WSe$_2$: 0.44–0.48 $m_e$)[15, 16], allowing high $I_{60}$ in BP TFETs (see equation (1)) as well as high carrier mobility[17,18]. These unique BP band properties enabled us to achieve the lowest SS$_{ave\_5dec} \approx 26$ mV/dec and the highest $I_{60} \approx 1.0$ µA/µm with $I_{60}/I_{off} \approx 3.6 \times 10^5$ among all TFETs reported thus far in our natural HJ-TFETs (NHJ-TFETs) with a single BP flake of varying thickness.

The single BP flake used in the NHJ-TFET consists of three distinct regions (Fig. 1a and Supplementary Fig. 8): (i) bulk BP (source), (ii) and (iii) ML BP not covered and covered by graphite/ultrathin hexagonal boron nitride (hBN) (channel and drain, respectively). Since the bulk BP band (region (i)) does not shift with gate voltage due to large thickness (60–100 nm) (Supplementary Fig. 9), both top-gate, $V_{TG}$, and back-gate voltage, $V_{BG}$ only affect ML BP band. While $V_{TG}$ shifts the ML BP band in region (ii), $V_{BG}$ shifts the ML BP band in both region (ii) and (iii). The use of ultrathin hBN (2–3 layers) between graphite and ML BP in region (iii) has advantages over direct graphite contact on ML BP (Supplementary Fig. 8); 1) Ultrathin hBN placed between graphite and ML BP protects ML BP region (iii) band structure, since strong chemical interaction between ML BP and metal atoms can destroy ML BP band structures and cause metallization[19]. 2) Fermi-level pinning does not occur in ML BP region (iii) since ultrathin hBN increases the distance between BP and the graphite layer. Fermi-level pinning is expected to occur due to chemical bonding between P and metal atoms regardless of the metal work function[19]. Therefore, ML BP region (iii) chemical potential can be tuned by the drain bias voltage applied to graphite $V_D$ with an ultrathin hBN placed between the graphite and ML BP (Supplementary Section 8). In addition, we fabricated devices such that carrier transport occurs in the armchair direction with smaller $m^*$ than in the zigzag direction to enhance the on-current.

Figure 1b,c shows representative transfer curves of two BP NHJ-TFET devices at $|V_D| \leq$ 0.7 V. Device 1 has bottom-gate dielectrics 285 nm $SiO_2$ and top-gate dielectrics 10 nm hBN while device 2 has bottom-gate dielectrics 3 nm hBN and top-gate dielectrics 5 nm hBN. The channel length ($L$) and width ($W$) are $L \approx 0.7$ μm, $W \approx 1$ μm in device 1 and $L \approx 0.5$ μm, $W \approx 1$ μm in device 2. Here, we demonstarte the first p-type TFET operation in device 1 with $SS_{ave\_4dec}$ < 60 mV/dec. Furthermore, device 1 p-type operation at $V_D$ = -0.6 V outperforms all previous TFETs (including n-type TFETs) for both $SS_{ave\_4dec}$ and $I_{60}$, with $I_{60}$ close to the preferred range (1–10 μA/μm)[2, 11]. The measured $I_D$ vs. $V_{TG}$ (Fig. 1b, gate dielectrics: 10 nm hBN) shows $SS_{ave\_4dec} \approx 23.7$ mV/dec and $I_{60} \approx 0.65$ μA/μm. The n-type operation in device 2 at $V_D$ = +0.7 V (Fig. 1c) shows ultralow $SS_{ave\_4dec}$ as well as the highest $I_{60}$ among all previous TFETs with sub-thermionic $SS_{ave\_4dec}$. The measured $I_D$ vs. $V_{BG}$ (gate dielectrics: 3 nm hBN) shows $SS_{ave\_4dec} \approx$ 24.0 mV/dec and $I_{60} \approx 0.054$ μA/μm. Note that the on/off switching requires much lower voltages $\Delta V_{TG}$ = 0.15 V and $\Delta V_{BG}$ = 0.2 V for devices 1 and 2 respectively than $\Delta V_G$ = 0.7 V in the state-of-the-art MOSFETs, implying much reduced power consumption in our BP NHJ-TFETs.

Device optimization for both n- and p-type TFETs is essential to develop low-power complementary TFET technology for beyond CMOS. As described above, positive (negative) $V_D$ shifts the region (iii) band downwards (upwards) (see the band diagrams in Fig. 1b,c). This control of region (iii) ML band edges by $V_D$ enables our BP devices to operate as complementary n- and p-type TFETs depending on the sign of $V_D$. Note that if $V_D$ did not shift the band of ML BP region (iii), our devices could not operate as a complementary TFET regardless of the sign of $V_D$ (Supplementary Section 8). At the on state, regions (ii) and (iii) ML BP bands are adjusted with gate voltages such that a tunneling energy window, $\Delta\phi > 0$ opens between the bulk and ML BP, and hence BTBT transmission probability (equation (1)) becomes significant. If the energy window is blocked, TFET becomes off state. For n-type TFET with $V_D > 0$, $\Delta\phi$ = [bulk BP valence band maximum] – [ML BP conduction band minimum]; whereas for p-type TFET with $V_D < 0$, $\Delta\phi$ = [ML BP valence band maximum] – [Bulk BP conduction band minimum].

We observed two on/off mechanisms in our heterostructured BP devices depending on $V_{BG}$ and $V_D$: thermal injection and BTBT as shown in Fig. 2a,b. Thermal injection occurred when barrier height is reduced inside the ML BP between region (ii) and (iii) by $V_{BG}$ so that carriers having higher energy than the barrier height according to the Boltzmann distribution can move over the barrier. On the other hand, BTBT occurred when the source (bulk BP, region (i)) and the channel (ML BP, region (ii)) became oppositely doped and the tunnel window $\Delta\phi > 0$ opened up (Supplementary Fig. 12). From the temperature dependent transfer curves, we extracted $SS_{ave\_4dec}$ and $SS_{ave\_2dec}$ respectively in BTBT and thermal injection regimes. The insets of Fig 2a,b show that both p- and n-type operations with $SS_{ave\_4dec} < 60$ mV/dec at $T = 300$ K retain constant $SS_{ave\_4dec}$ across 8–300 K, confirming the carrier injection mechanism is BTBT. When BTBT dominates, thermally activated sections of the source Fermi distributions below conduction band minimum and above valence band maximum of the source and channel are effectively blocked and hence tunneling probability (equation (1)) is temperature independent. Therefore, the electronic system is effectively cooled, i.e., maintains low temperature. On the other hand, SS shows linear dependence on $T$ in the thermal injection limit due to the exponential increase in the transport of thermally activated carriers over potential barrier with $T$. Note that the temperature-dependent transfer curves show an additional feature in BTBT transitions; The BTBT threshold voltage shifts with temperature (Fig. 2 and Supplementary Fig. 13). We consider a few possibilities to explain BTBT threshold voltage shifts. Most of previous experimental and theoretical studies have shown that threshold voltage shifts of TFETs with temperature originate from trap-assisted-tunneling (TAT), Shockley-Read-Hall (SRH) recombination, or Urbach tail (exponentially decaying evanescent states near the band edges). However, TAT, SRH, or Urbach tail were the major roadblocks to improving since they increase both SS and $I_{off}$ with temperature[20], inconsistent with our observation of constant SS and $I_{off}$. Another possibility is that ML BP bandgap reduction with temperature could induce threshold voltage shift (Supplementary Fig. 14). In this case, BTBT on-current also increases with temperature due to the decrease in ML BP bandgap and thereby increase in BTBT energy window between the bulk and ML BP with temperature. While only thermal injection contributes to the electron side IV curve in Fig. 2a, both thermal injection and BTBT inside ML BP contributes to the hole side IV curves in Fig. 2b. Since the thermal injection, i.e., the mechanism of carrier transport in conventional CMOS transistor, dominates the transport in the electron transport in Fig. 2a, on-current should decrease with temperature due to phonon scattering and thereby mobility degradation. Since both thermal injection and BTBT inside ML BP contribute to the hole side IV curves in Fig. 2b and BTBT on-current significantly depends on the BTBT energy window, which increases with temperature due to the decrease in the ML BP bandgap, on-current increases with temperature in the hole transport at positive $V_D$ in Fig. 2b oppositely to the electron transport at negative $V_D$ in Fig. 2a. The on/off transition of electron transport dominated by thermal injection in Fig. 2a occurs at the same back-gate voltage with changing temperature, implying that the conduction band edge of drain ML BP does not change with temperature, which might be related to the schottky barrier formed between graphite contact and ML BP[21]. The decrease in the bandgap of ML 2D van der Waals layers has been observed in transition metal dichalcogenides such as ML $MoS_2$[22]. To our best knowledge, there have been neither experimental nor theoretical reports on the temperature dependence of ML BP bandgap. Another possibility of temperature-dependent threshold voltage shift is work function change of back and top gate with temperature. According to simulation studies[23-25], work function change of gate in our devices can induce threshold voltage shift with constant SS and $I_{off}$. To determine the accurate mechanism of temperature-dependent threshold voltage shift in our BP NHJ-TFETs requires further systematic studies in the future.

Figure 3 shows comparison of BP NHJ-TFETs with the only two previously reported n-type TFETs to achieve $SS_{ave\_4dec}$ < 60 mV/dec and the state-of-the-art intel 14 nm Si MOSFET[26]. We extracted $I_D$ vs. SS and $I_{60}$ vs. $SS_{ave\_4dec}$ from transfer curves shown in Supplementary Fig. 20. The comparison data clearly indicate that 1) BP NHJ-TFETs outperform all previous TFETs and 2) more importantly, BP NHJ-TFETs for the first time fullfill the two critical requirements simultaneously (sub-thermionic $SS_{ave\_4dec}$ and high $I_{60}$) in both p and n-type operations. To date, only two n-type and no p-type TFETs have been reported to achieve a sub-thermionic $SS_{ave}$ < 60 mV/dec over 4 decades of current, but with 2–5 orders of magnitude lower $I_{60}$ than the required range, 1–10 µA/µm. Figure 3a shows that the BP NHJ-TFETs exhibit 1) 3–4 orders of magnitude lower $I_{off}$ than the intel 14nm Si MOSFET (Supplementary Fig. 20), indicating standby power consumption is reduced by $10^3$–$10^4$, and 2) $I_{60}$ was demonstrated up to ≈ 1 μA/μm comparable to the on-current near the threshold voltage in MOSFETs, suggesting these NHJ-TFETS would be competitive replacement for low-power switches operating at fast enough rates. Figure 3b clearly shows that BP NHJ-TFET performance is the closest to the preferred region, with the two major figures of merit, i.e., low $SS_{ave\_4dec}$ and high $I_{60}$, among all TFETs. In addition, Supplementary Table 1 compares other previous TFETs and for $SS_{ave\_5dec}$ and $I_{60}/I_{off}$.

Although the lowest drain voltage we used was $V_D$ = -0.6 V for p-type and $V_D$ = +0.7 V for n-type operation, $V_D$ is expected to be further reduced with enhanced performance if we use high-κ dielectric materials between graphite and ML BP, or control chemical doping of the drain region. This could also improve the on-current of the BP NHJ-TFETs to that for MOFET (100–1000 μA/μm) at low bias ≤ 0.5V owing to the absence of interface problems in the natural BP HJ[27].

**References (in main text)**

## Acknowledgments

We thank P. Kim, A. Seabaugh, R. Sajjad, E. Yablonovitch, F. Liu, G. Klimeck, H.J. Choi, E.H. Hwang for helpful discussions. We also thank C. Lee for help with the dry-transfer technique. S. Cho acknowledges support from Korea NRF (Grant No. 2019M3F3A1A03079760 and Grant No. 2016R1A5A1008184), Samsung Electronics Device Solution Basic Science Program, and KI 2019 Transdisciplinary Research Program. K.W. and T.T. acknowledge support from the Elemental Strategy Initiative conducted by the MEXT, Japan, A3 Foresight by JSPS and the CREST (JPMJCR15F3), JST.

## Author contributions

S. Cho conceived and supervised the project. S.K. fabricated devices and performed measurements. G.M. and W.S. assisted fabrication of devices. G.M. and S. Chang assisted Raman and photoluminescence measurement of BP flakes. K.W. and T.T. grew high-quality hBN single crystals. G.M., W.S., H.L., B.K. and T.J. assisted low-temperature transport measurements. S. Cho and S.K. analyzed the data and wrote the manuscript. All the authors contribute to editing the manuscript.

## Competing interests

The authors declare no competing financial interests.

## Additional information

**Supplementary information** is available in the online version of the paper.

**Reprints and permissions information** is available at www.nature.com/reprints.

**Correspondence and requests for materials** should be addressed to S. Cho.

# Methods

**Device fabrication and measurement.** We first prepared graphite, hBN, and BP flakes on 90nm $SiO_2$ wafer by mechanical exfoliation from bulk crystals in an argon-filled glove box (< 0.1 ppm of $H_2O$ and $O_2$) to maintain clean surfaces and prevent contamination from air exposure. Subsequently, by using a polycarbonate (PC)-film-covered polydimethylsiloxane (PDMS) stamp[28], each flake was picked up from the substrate at 80°C and eventually released on a 285nm $SiO_2$/highly doped Si substrate at an elevated temperature. The PC film was washed out by chloroform followed by acetone and IPA. Then, we used standard e-beam lithography and $CF_4$ plasma etching followed by e-beam evaporation to prepare electrical contacts to the van der Waals layers. Finally, additional metal deposition was performed to form the top-gate. After the device fabrication is complete, we performed DC measurement to obtain the transfer curves at temperatures from 300K to 8K in a home-built cryostat. We used Keithley 2400 or Yokogawa 7651 to bias DC voltages to the drain (or the gate) electrodes. To measure the drain currents, we used a preamplifier DDPCA-300 to amplify the drain current signal by $10^6$ and convert it to a voltage and measured the amplified voltage signal by using a Keithley 2182a nanovoltmeter. To increase accuracy and reduce noise level, we averaged 30 times for each data points taken at every 5 mV of gate steps both for upward and downward gate sweeps. Each transfer curve ($I_D$ vs. $V_G$) plots absolute values of $I_D$ in log scale as a function of $V_G$.

**ML BP characterization.** We have identified ML BP by combining three different characterization methods: atomic force microscopy (AFM), Raman spectroscopy, and photoluminescence (PL) (Supplementary Section 5). In addition, we have performed polarized-Raman spectroscopy to determine the crystal orientation of BP flakes. We have determined the armchair direction along which $A_g^1$ and $A_g^2$ modes show maximum peak intensities while $B_{2g}$ mode shows minimum peak intensity[29, 30] (see Supplementary Section 4).

# Data availability

The data that support the findings of this study are available from the corresponding authors upon reasonable request.

**References (in Methods)**

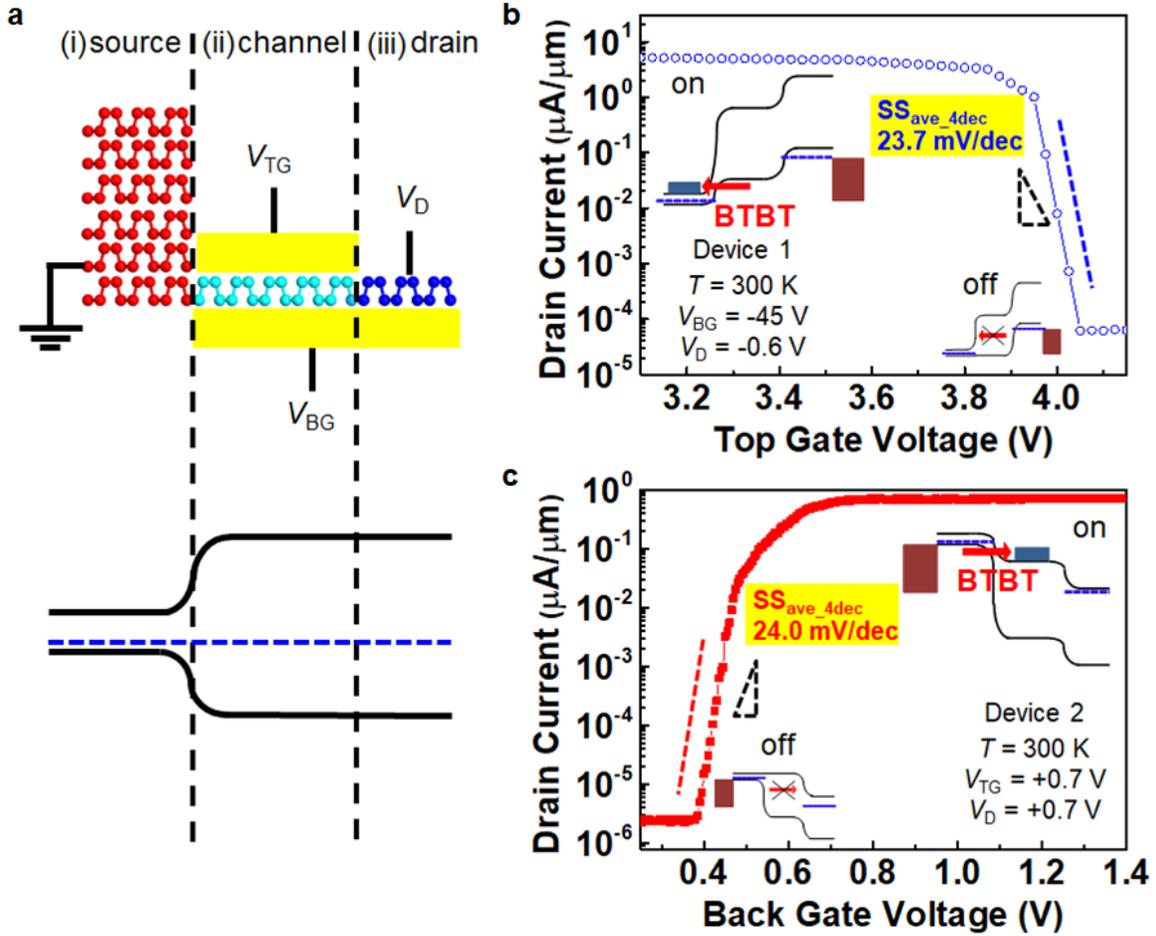

**Figure 1. Black phosphorous (BP) band properties and transfer curves for two fabricated BP natural heterojunction-tunnel field effect transistors (NHJ-TFETs) at $|V_D| \leq 0.7$ V. a,** BP NHJ-TFET schematic structure and BP band diagram in each part of source, channel, and drain. The differently-colored BP in each region represents different doping level of each region. Operation type is determined by the drain bias voltage ($V_D$) sign, which shifts the ML BP (region (iii)) band. **b,** P-type transfer curve for device 1 at $V_D$ = -0.6 V. On/off switching requires $\Delta V_{TG}$ = 0.15 V. **c,** N-type transfer curve for device 2 at $V_D$ = +0.7 V. On/off switching requires $\Delta V_{BG}$ = 0.2 V. Blue and red dotted lines in **(b)** and **(c)** represent subthreshold slopes for the device 1 and 2 respectively, while the dotted triangles show subthreshold slopes of SS = 60 mV/dec for

comparison. (Inset) By controlling $V_{BG}$, $V_{TG}$, and $V_D$, ML BP (region (ii) and (iii)) band can be adjusted such that a tunneling energy window ($\Delta\phi > 0$) occurs between bulk and ML BP, enabling band-to-band tunneling (BTBT). If the energy window is blocked, the TFET becomes off state.

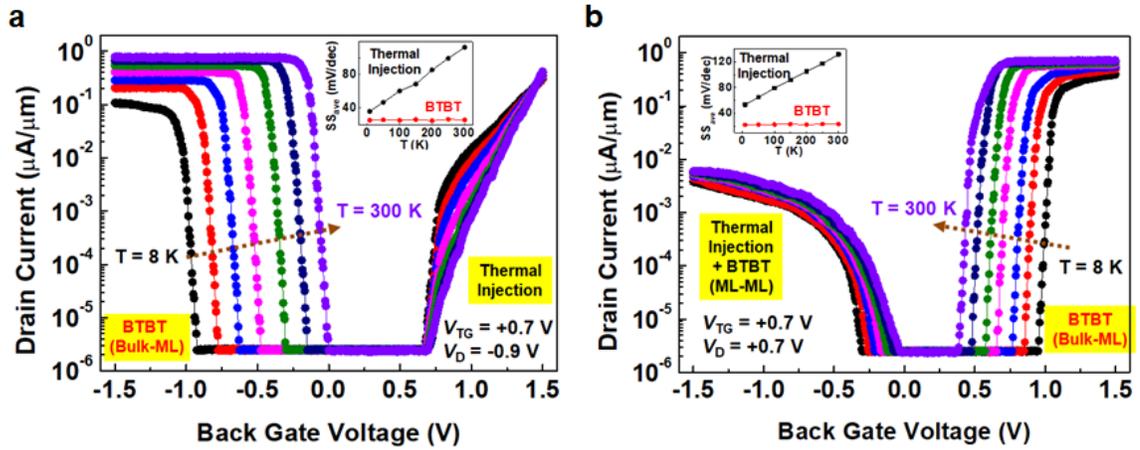

**Figure 2. Temperature-dependent transfer curves originating from two different carrier injection mechanisms: Band-to-band tunneling (BTBT) and thermal injection.** Temperature-dependent transfer curves for **(a)** p-type and **(b)** n-type NHJ-TFET operation, respectively. From black to purple curve, we show transfer curves measured at temperature from 8K to 300K with 50K steps. BTBT threshold voltage shifts with temperature. This might be related to the increase in bandgap of ML BP at lower temperature. (Inset) Black lines and dots represent $SS_{ave\_2dec}$ measured in the thermal injection limit, changing linearly with temperature. Red lines and dots represent $SS_{ave\_4dec}$ measured in BTBT limits, which is almost independent of temperature. Error bars represent the standard deviations of $SS_{ave}$.

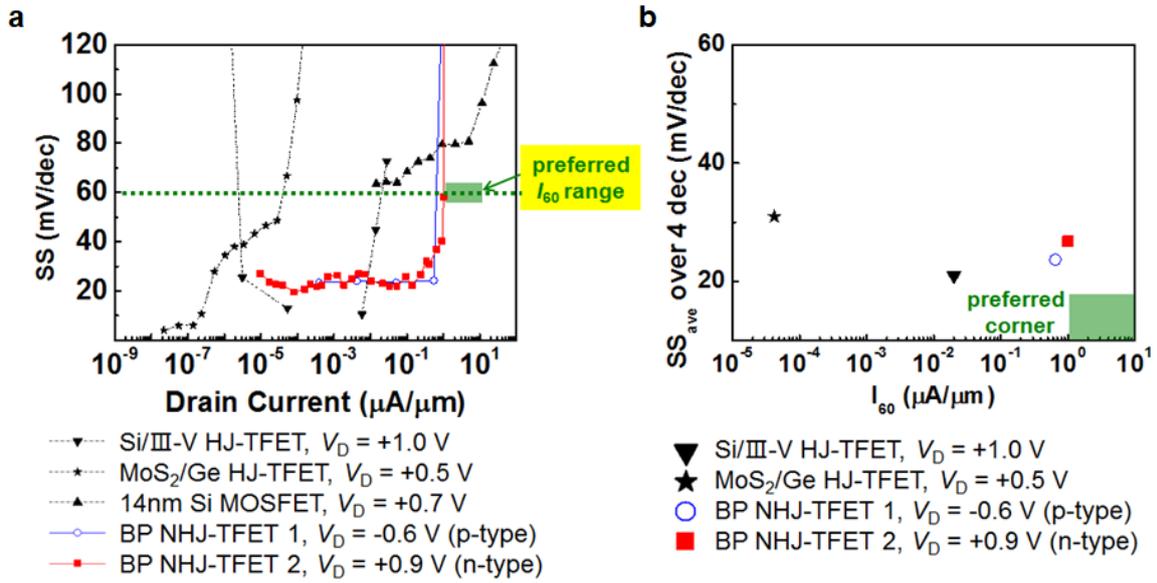

**Figure 3. Performance comparison for the black phosphorus natural heterojunction-tunnel field effect transistors (BP NHJ-TFETs) with previous n-type TFETs reporting sub-thermionic subthreshold swing average over 4 decades ($SS_{ave\_4dec}$) and state-of-the-art intel 14 nm Si MOSFET. a,** $I_D$ vs. SS data extracted from transfer curves in Supplementary Fig. 20. Olive dotted line represents SS = 60 mV/dec and olive square represents the preferred $I_{60}$ range (1–10 μA/μm). **b,** $I_{60}$ vs. $SS_{ave\_4dec}$ calculated from transfer curves in Supplementary Fig. 20. Olive square represents the preferred $I_{60}$ range of 1–10 μA/μm with low $SS_{ave\_4dec}$ < 20 mV/dec. Black inverted triangle, star, and triangle represent Si/III–V HJ-TFET, MoS$_2$/Ge HJ-TFET, and n-type 14 nm Si MOSFETs, respectively. Blue hollow circle represents p-type BP NHJ-TFET 1 at $V_D$ = -0.6 V and red square represents n-type BP NHJ-TFET 2 at $V_D$ = +0.9 V.